\def\draftversion{false}

\RequirePackage{ifthen}
\ifthenelse{\equal{\draftversion}{true}}{
  \documentclass[aps,prb,10pt,galley,amsmath,amssymb,showpacs,
                 superscriptaddress]{revtex4}
}{
  \documentclass[aps,prb,10pt,twocolumn,amsmath,amssymb,showpacs,
                 superscriptaddress]{revtex4-1}
}

\usepackage{graphicx}
\usepackage[usenames]{color} 
\usepackage{bm} 
\usepackage{dcolumn} 
\usepackage{soul} 
\usepackage{multirow}

\ifthenelse{\equal{\draftversion}{true}}{
  \marginparwidth 2.7in
  \marginparsep 0.5in
  \newcounter{comm} 
  \def\commnext{\stepcounter{comm}}
  \def\commtext{{\bf\color{blue}[\arabic{comm}]}}
  \def\commmar{{\bf\color{blue}[\arabic{comm}]}}
  \def\dvm#1{\commnext\marginpar{\small DV\commmar: #1}\commtext}
  \def\mym#1{\commnext\marginpar{\small MY\commmar: #1}\commtext}
  \def\mlab#1{\marginpar{\small\bf #1}}
  
}{
  \def\dvm#1{}
  \def\mym#1{}
  \def\mlab#1{}
  
}

\def\beq{\begin{equation}}
\def\eeq{\end{equation}}

\def\XO{$X_2$O$_3$}

\def\ABO{$AB$O$_3$}
\def\ABB'O{$A_2BB'$O$_6$}
\def\AA'BB'O{$AA'BB'$O$_6$}
\def\BTO{BaTiO$_3$}
\def\LNO{LiNbO$_3$}
\def\LTO{LiTaO$_3$}
\def\ZSO{ZnSnO$_3$}
\def\MTO{MnTiO$_3$}
\def\FTO{FeTiO$_3$}
\def\LZTO{Li$_2$ZrTeO$_6$}
\def\LHTO{Li$_2$HfTeO$_6$}
\def\MFWO{Mn$_2$FeWO$_6$}
\def\MWO{Mn$_3$WO$_6$}
\def\ZFOO{Zn$_2$FeOsO$_6$}

\begin{document}


\title{Ferroelectricity in corundum derivatives}

\author{Meng Ye}
\email{mengye@physics.rutgers.edu}
\affiliation{
Department of Physics \& Astronomy, Rutgers University,
Piscataway, New Jersey 08854, USA}

\author{David Vanderbilt}
\affiliation{
Department of Physics \& Astronomy, Rutgers University,
Piscataway, New Jersey 08854, USA}

\date{\today}

\begin{abstract}
The search for new ferroelectric (FE) materials holds promise for
broadening our understanding of FE mechanisms and extending the range
of application of FE materials. Here we investigate a class of \ABO\
and \ABB'O materials that can be derived from the \XO\ corundum
structure by mixing two or three ordered cations on the $X$ site.
Most such corundum derivatives have a polar structure, but it is
unclear whether the polarization is reversible, which is a requirement
for a FE material. In this paper, we propose a method to study the FE
reversal path of materials in the corundum derivative family. We first
categorize the corundum derivatives into four classes and show that
only two of these allow for the possibility of FE reversal. We then
calculate the energy profile and energy barrier of the FE reversal
path using first-principles density functional methods with a structural
constraint. Furthermore, we identify several empirical measures that
can provide a rule of thumb for estimating the energy barriers.
Finally, the conditions under which the magnetic ordering is compatible
with ferroelectricity are determined. These results lead us to predict
several potentially new FE materials.
\end{abstract}

\maketitle


\section{Introduction}
Ferroelectricity requires a material to have a spontaneous electric
polarization that can be reversed by an external electric field.
\cite{BookRabe} The search for new ferroelectric (FE) materials holds
promise for broadening our understanding of FE mechanisms and extending
the range of application of FE materials. A switchable spontaneous
polarization implies a hysteresis effect that can be used in memory
devices. \cite{BookScott} FE materials also exhibit high and tunable
electric permittivity, which can be used in capacitors to increase the
capacitance and reduce the size of devices. In addition, FE materials
are piezoelectric and pyroelectric, according to symmetry considerations.
These combined properties make FE materials ideal for electric,
mechanical and thermal sensors. Recently, research on multiferroics,
in which FE and ferromagnetic orders coexist in the same material, has
further extended the range of application of FE materials.
\cite{Scott06,Cheong&Mostovoy07,Ramesh&Spaldin07,Wang09,Khomskii09,
Picozzi15,Ghosez15}

The most intensively studied family of FE oxides is that of the
perovskite oxides such as \BTO. \cite{pvskFE92,pvskFE97} Perovskite
oxides have the chemical formula \ABO\ with the $A$ cation much
larger than the $B$ cation. The FE distortion is usually driven by
$B$-site off-centering and typically requires an empty $d$ shell on
the $B$ cation, which is not compatible with magnetism. Recently,
rocksalt-ordered \ABB'O\ double perovskites (and more complex \AA'BB'O
materials) have also attracted great interest. \cite{dpvskFE13, dpvskMF14}

The corundum derivatives \ABO\ and \ABB'O\ make up a family of oxides
that can be derived from the corundum structure with cation ordering.
Most corundum derivatives are polar and thus can potentially be new
FE oxides. \cite{CuNb/TaO, Fe/Mn/NiTiO, FeTiO, ZnSnOFE,CdPbO/PbNiO,
MnTi/SnO,PbNiO,ScFeO,LiZr/HfTeO,Mn2FeSbO,Mn2FeNb/TaO,Mn2FeMoO,
Zn2FeTaO,Mn2FeWO,Zn2FeOsO,Mn3WO} \LNO\ (LNO) is a well-known example
of a FE corundum derivative. \cite{LNO/LTO, LNO/LTOcal, LNO/LTOcal2}
Despite the similar chemical formula, corundum derivatives are
different from perovskites in many aspects.
The high-symmetry parent structure is rhombohedral for corundum
derivatives but cubic for perovskites.
The polarization reversal mechanisms are also distinct.
In corundum derivatives, the polarization reversal is driven by the
small $A$ or $B$ cations migrating between oxygen octahedra, \cite{LNO/LTOcal,LNO/LTOcal2}
so that $d^{\,0}$ configuration is not required.
This is in contrast to the off-centering displacement of $d^{\,0}$
$B$ cations in the oxygen octahedra in most perovskites.
The huge number of potential
combinations of $A$, $B$ and $B'$ cations in the corundum-derivative
family opens the possibility to achieve not only
ferroelectricity but also multiferroicity.  \cite{Fe/Mn/NiTiO,
FeTiO}

In this paper, we use first-principles density functional methods to
systematically study the polar structure and the coherent FE reversal
paths for a variety of corundum derivatives. First, the structures
of corundum derivatives are introduced in Sec.~\ref{sec:structure}.
Then the structural criteria for corundum derivatives to be FE are
discussed in Sec.~\ref{sec:criteria}. In Sec.~\ref{sec:profile}, we
propose a systematic method to analyze the coherent FE barrier and
energy profile. Lastly, the new method is applied to several corundum
derivatives, and empirical measures that can provide a rule of thumb
for estimating the reversal barrier are summarized in Sec.~\ref{sec:results}.

\section{Preliminary}
\subsection{Structure}\label{sec:structure}
\begin{figure}
  \includegraphics[width=8cm]{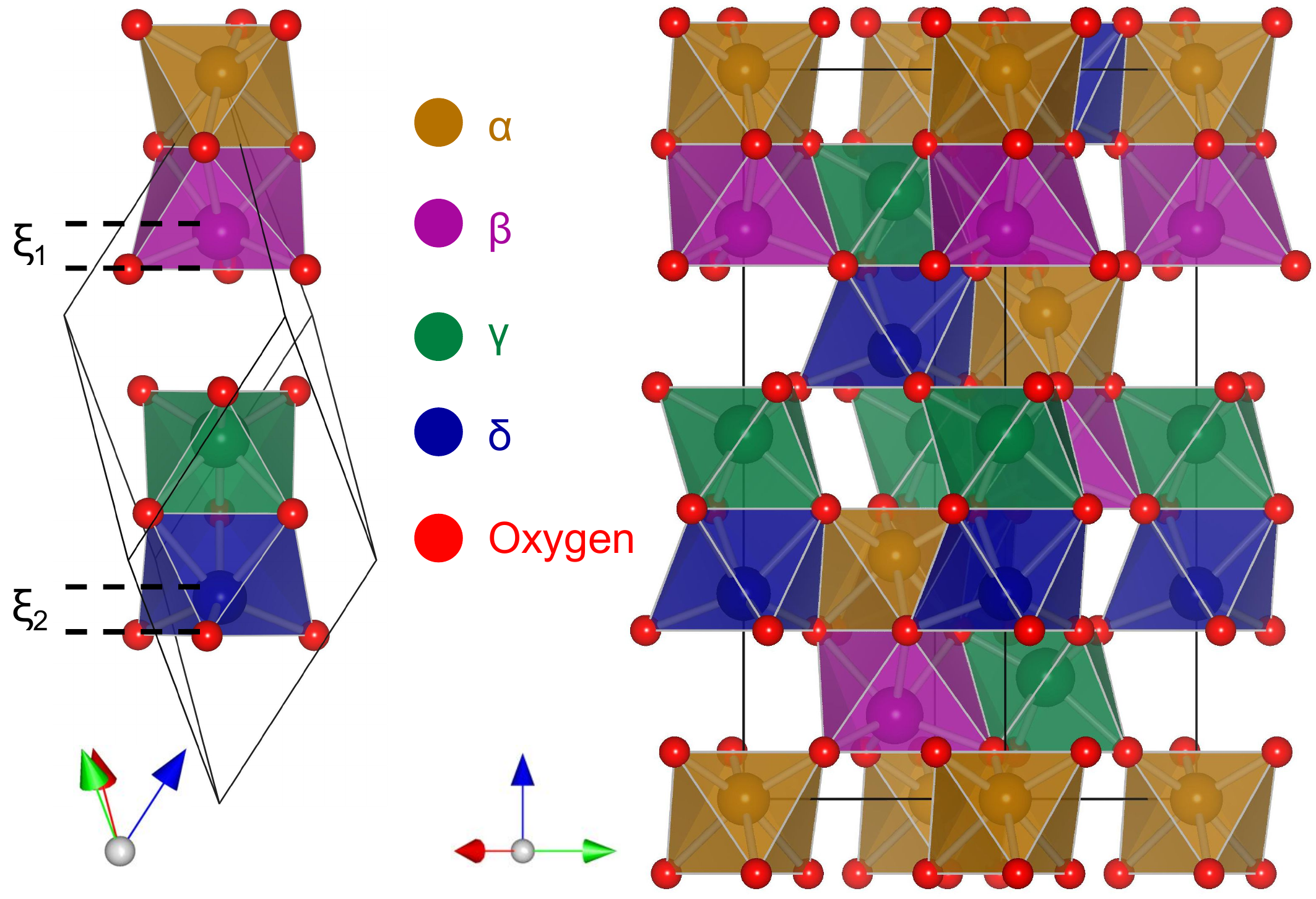}
    \caption{Structure of corundum derivatives. The unit cell in
    the rhombohedral setting is shown at the left; an enlarged
    hexagonal-setting view is shown at right.
    The cations $\alpha$, $\beta$, $\gamma$, and $\delta$ are
    are all identical in the \XO\ corundum structure.
    For the LNO-type \ABO, $\beta=\delta=A$, $\alpha=\gamma=B$;
    for the ilmenite \ABO, $\beta=\gamma=A$, $\alpha=\delta=B$;
    for the ordered-LNO \ABB'O, $\beta=\delta=A$, $\gamma=B$, $\alpha=B'$;
    for the ordered-ilmenite \ABB'O, $\beta=\gamma=A$, $\delta=B$, $\alpha=B'$.
    At left, $\xi_1$ (or $\xi_2$) is the distance between $\beta$ (or $\delta$)
    and the oxygen plane that it penetrates during the polarization reversal.}
    \label{fig:structure}
\end{figure}
%
The corundum derivatives \ABO\ and \ABB'O\ can be derived from the
corundum structure \XO\ with cation ordering as shown in Fig.~\ref{fig:structure}.
In the 10-atom rhombohedral unit cell, the cations are spaced along
the three-fold rotation axis and each one is surrounded by a distorted
oxygen octahedron. Two thirds of the oxygen octahedra are filled with
cations, while one third of them are cation-vacant.

Based on the combinations and arrangements of cations, the corundum
derivatives can be classified into four types, which we denote as
follows. An oxygen octahedron with an $A$ or $B$ cation inside is
written as ``$A$" or ``$B$", and if an oxygen octahedron is cation-vacant,
the octahedron is denoted by ``$-$". This notation is then used
to represent the column of six face-sharing oxygen octahedra in the
unit cell for
each of the four different types of corundum derivatives.  Thus,
the LNO-type \ABO\ is $AB$$-$$AB$$-$, the ilmenite \ABO\ is $AB$$-$$BA$$-$, the
ordered-LNO \ABB'O\ is $AB$$-$$AB'$$-$, and the ordered-ilmenite \ABB'O\ is
$AB$$-$$B'A$$-$. Other combinations, such as $AA$$-$$BB$$-$, are
connected to the four existing types
as explained in Sec.~\ref{sec:criteria}.

\subsection{Coherent FE polarization reversal}\label{sec:criteria}

Experimentally, the FE polarization reversal process is a complicated
one that typically proceeds by nucleation and motion of domain walls,
involving both intrinsic atomic motion and extrinsic pinning by defects.
In this paper we have chosen to focus only on coherent FE domain
reversal, in which every unit cell undergoes the polarization reversal
simultaneously. This coherent reversal process is clearly highly
oversimplified, but previous studies of perovskite oxides have shown
that the coherent barrier provides a figure of merit that is often a
useful indicator of the real barrier to polarization reversal.

The atomic origin of ferroelectricity in LNO is well-understood.
\cite{LNO/LTOcal, LNO/LTOcal2} In LNO, the polarization reversal is
driven by an infrared-active (IR-active) mode that is associated with
the motion of Li cations along the rhombohedral axis. In the reversal
process, each Li cation penetrates through an oxygen plane and migrates
into an adjacent unoccupied oxygen octahedron. In our notation, the
polarization reversal process interchanges Li with $-$, so that the
polar structure changes from LiNb$-$LiNb$-$ to its inversion image
$-$NbLi$-$NbLi.

For all types of polar corundum derivatives, we assume that the
polarization reversal mechanism is similar to that of LNO, i.e., that
the reversal
process interchanges $A$ or $B$ with $-$. The structures before and
after this process are listed in Table~\ref{tab:FEswitch}.
Under such
an operation, the LNO-type structure is transformed into its own
inversion-reversed image, which is a typical FE behavior.
The same is true for the ordered-LNO structure.  By contrast,
the ilmenite-type $AB$$-$$BA$$-$ is transformed into $BB$$-$$AA$$-$, and
the ordered-ilmenite $AB$$-$$B'A$$-$ into $BB'$$-$$AA$$-$.
These structures are not inversion-equivalent to the starting
structures.  Moreover, they exhibit face-sharing $A$-containing
octahedra, making them relatively unfavorable energetically.
Additionally, both the ilmenite structure and its switched partner
are centrosymmetric.  For these reasons, we exclude the ilmenite
and ordered-ilmenite structures from further consideration as FE
candidates.

\begin{table}
\begin{center}
\begin{ruledtabular}
    \caption{Corundum-derived structures before and after
    polarization reversal.}
    \begin{tabular}{lll}
    &\multicolumn{1}{c}{Before} &\multicolumn{1}{c}{After}\\ \hline
    LNO$$-$$type       & $AB$$-$$AB$$-$  & $-$$BA$$-$$BA$ \\
    Ilmenite        & $AB$$-$$BA$$-$  & $-$$BB$$-$$AA$ \\
    Ordered-LNO     & $AB$$-$$AB'$$-$ & $-$$BA$$-$$B'A$\\
    Ordered ilmenite & $AB$$-$$B'A$$-$ & $-$$BB'$$-$$AA$\\
    \end{tabular}\label{tab:FEswitch}
\end{ruledtabular}
\end{center}
\end{table}

\subsection{Energy profile calculations}\label{sec:profile}
For the LNO-type and the ordered-LNO FE candidates, we firstly analyze
the symmetry of the reversal path. The ground state symmetry is R3c
for the LNO-type materials, and R3 for the ordered-LNO ones,
but the symmetry of the
reversal path is not straightforward. In this paper, we assume that the
three-fold rotation is always preserved, so that the symmetry of the
path can only be R3c or R3 for the LNO-type case, and R3 for the ordered-LNO
case.  If the structure acquires an inversion center at the midpoint
of the path when the polarization is zero (R$\bar 3$c or R$\bar 3$
for the two cases respectively), the energy profile would be symmetric.
If the inversion symmetry at the midpoint is broken, as for example by
magnetic ordering, the energy profile would be asymmetric.

Based on the symmetry of the reversal path, we adopt different
methods to calculate the energy profile of the FE reversal. In the case
when the inversion symmetry is present at the midpoint structure, the
polarization reversal can be analyzed in terms of an unstable IR-active
phonon mode at the high-symmetry midpoint. In general, even if the
midpoint is not in a high-symmetry reference structure, the motion of
the small $A$ cations is responsible for the polarization reversal.
As illustrated in Fig.~\ref{fig:structure}, we define $\xi_1$ ($\xi_2$)
to be the distance between the first (second) $A$ cation and the oxygen
plane that it penetrates as this $A$ cation moves along its path. Then
$\xi_1 + \xi_2$ is adopted as a ``reaction coordinate'' to describe the
reversal. Finally, we use either the unstable IR-active mode at the
midpoint (for the high-symmetry case) or  $\xi_1 + \xi_2$ (otherwise)
as a structural constraint, and relax all other internal structural
degrees of freedom while stepping through a sequence of values of this
constraint. This gives us the energy profile along the path, from which
the energy barrier is obtained by inspection.

\subsection{First-principles methods}\label{sec:fpm}
Our calculations are performed with plane-wave density functional
theory (DFT) implemented in VASP. \cite{VASP} The exchange-correlation
functional that we use is PBEsol, a revised Perdew-Burke-Ernzerhof
generalized-gradient approximation that improves equilibrium properties
of densely-packed solids. \cite{PBEsol} The ionic core environment is
simulated by projector augmented-wave (PAW) pseudopotentials. \cite{PAW}
For transition metal elements Mn and Fe, we use a Hubbard $U=4.2$\,eV
on the 3$d$ orbitals. \cite{DFTU,Mn2FeMoO} For the Os 5$d$ orbital, we
use a Hubbard $U=1.4$\,eV. \cite{Zn2FeOsO} The magnetic moments are
collinear and spin-orbit coupling is neglected.
The cutoff energy for all calculations is 550\,eV.
The energy error threshold varied slightly in different calculations,
but an accuracy between $1.0\times10^{-5}$ and $1.0\times10^{-7}$\,eV
is achieved in all calculations. The forces are reduced below
0.001 eV/Angstrom for calculations of structural relaxation.
A $6\times6\times6$ Monkhorst-Pack k-mesh is used in the calculations.
Linear-response methods are used to calculate the $\Gamma$-point
force-constant matrices. The spontaneous polarization is calculated
using the Berry phase formalism. \cite{ModernPolar}

\section{Results and discussion}\label{sec:results}
In this paper, we apply the method of calculating the energy
profile described in Sec.~\ref{sec:profile} to fully analyze the
coherent FE reversal path of the LNO-type corundum derivatives
\LNO,\cite{LNO/LTO} \LTO,\cite{LNO/LTO} \ZSO,\cite{ZnSnOFE}
\FTO,\cite{Fe/Mn/NiTiO,FeTiO} and \MTO,\cite{Fe/Mn/NiTiO} and the
ordered-LNO corundum derivatives \LZTO,\cite{LiZr/HfTeO}
\LHTO,\cite{LiZr/HfTeO} \MFWO,\cite{Mn2FeWO} \MWO,\cite{Mn3WO} and
\ZFOO.\cite{Zn2FeOsO}

\subsection{Ground state structure and magnetic order}
The properties of FE materials are sensitive to atomic displacements
and strain, so it is essential to start our calculation with accurate
structural parameters. The lattice constants and Wyckoff positions
obtained from our calculations are summarized in the Supplement, with
experimental results provided for reference. Our structural parameters
are very close to the experimental results. The oxidation states,
obtained by rounding the integrated charge around each cation, are
also displayed in Table~\ref{tab:oxidation} and are in good agreement
with experimental observations.
\begin{table}
\begin{center}
\begin{ruledtabular}
    \caption{Oxidation states of the LNO-type \ABO\ and the ordered-LNO
    \ABB'O corundum derivatives. The oxidation state of O ion is $-$2 in
    all materials.}
    \begin{tabular}{lccclccc}
    \multicolumn{1}{c}{LNO-type}&\multicolumn{1}{c}{$A$} &\multicolumn{1}{c}{$B$} &&
    \multicolumn{1}{c}{Ordered LNO}&\multicolumn{1}{c}{$A$} &\multicolumn{1}{c}{$B$} &\multicolumn{1}{c}{$B'$}\\ \hline
    \LNO  &+1 &+5 &&\LZTO &+1 &+4 &+6 \\
    \LTO  &+1 &+5 &&\LHTO &+1 &+4 &+6 \\
    \ZSO  &+2 &+4 &&\MFWO &+2 &+2 &+6 \\
    \FTO  &+2 &+4 &&\MWO  &+2 &+2 &+6 \\
    \MTO  &+2 &+4 &&\ZFOO &+2 &+3 &+5 \\
    \end{tabular}\label{tab:oxidation}
\end{ruledtabular}
\end{center}
\end{table}

The on-site magnetic moments are investigated for \FTO, \MTO, \MFWO,
\MWO\, and \ZFOO. Our DFT+$U$ calculation predict that the magnetic
moment is about 3.7~$\mu_B$ on each Fe$^{2+}$, 4.6~$\mu_B$ on each
Mn$^{2+}$, and 4.2~$\mu_B$ on Fe$^{3+}$. These results are in agreement
with the $d^6$ state of Fe$^{2+}$ and the $d^5$ configuration of Fe$^{3+}$
and Mn$^{2+}$. The magnetic moment on Os$^{5+}$ is
2.1~$\mu_B$ from our calculation, which is consistent with the high-spin
$d^3$ state after taking into account the screening of the Os moment
arising from the hybridization between Os 5$d$ and O 2$p$ orbitals.

The energy of different magnetic orderings is also studied. In our
calculation, we only consider magnetic structures that preserve the
periodicity of the rhombohedral unit cell. Our results suggest that
the ground-state magnetic ordering is anti-ferromagnetic (AFM) for
\FTO\ and \MTO\ and ferrimagnetic (FiM) for \ZFOO. To investigate the
magnetic structures of \MFWO\ and \MWO, four different types of unit
cell are considered in the calculation. We adopt a notation like
``$udu$" to describe the possible spin structure, where ``$u$" is
spin-up, ``$d$" is spin-down, and the spins are given on atom $\beta$,
$\delta$ and $\gamma$, in that order. The four possible states (not
counting those that are trivially related by a global spin reversal)
are $uuu$, $uud$, $udu$, and $udd$. The energy of each fully-relaxed
magnetic structure is listed in the Supplement. Of those, the most
stable state for both \MFWO\ and \MWO\ is $udu$. However, for \MWO,
the energy difference between the $uud$ and $udu$ states is tiny, so
we considered the polarization reversal for both magnetic states.

\subsection{Symmetry of the reversal path}\label{sec:sympath}

For the LNO-type materials, the simplest possible reversal path would
be one in which the two $A$ cations move synchronously, so that $\xi_1=\xi_2$
everywhere along the path. In this case the symmetry along the path is
R3c, except at the midpoint where there is an inversion center and the
symmetry becomes R$\bar3$c. Another possibility is that the cations
move sequentially, one after the other, so that $\xi_1\ne\xi_2$ for at
least part of the path. In this case the symmetry is R3 except at the
R$\bar3$ midpoint. In order to find out which scenario occurs, we
calculate the energy of the midpoint structures with symmetry R$\bar3$c
and R$\bar3$ respectively; the results are shown in Table \ref{tab:midpoint}.
For all LNO-type materials that we have studied, the R$\bar3$ midpoint
structure is energetically favored, which implies that the reversal
occurs via the lower-symmetry R3$\rightarrow$R$\bar3$$\rightarrow$R3
scenario, at least in the central portion of the path. This striking
result demonstrates that the midpoint of the FE reversal path in the
LNO-type FE materials is \textit{not}
identified with the high-temperature paraelectric structure,
\cite{LNO/LTOcal,LNO/LTOcal2} but instead has lower symmetry.

\begin{table}
\begin{center}
\begin{ruledtabular}
    \caption{The energy and the unstable phonon modes at the midpoint
    structure of \LNO, \LTO, \ZSO, \FTO\ and \MTO\ with symmetry
    R$\bar3$c and R$\bar3$.
    The energy of the ground-state structure is set to be zero as
    reference and the unit is meV per unit cell. The imaginary frequency of the
    unstable phonon is given in units of cm$^{-1}$.}
    \begin{tabular}{ccccccc}
    &\multicolumn{2}{c}{Energy} &&\multicolumn{3}{c}{Unstable modes}\\
    &\multicolumn{1}{c}{R$\bar3$c}&\multicolumn{1}{c}{R$\bar3$} &&\multicolumn{2}{c}{R$\bar3$c}&\multicolumn{1}{c}{R$\bar3$}\\
    &&&&\multicolumn{1}{c}{A$_{2u}$}&\multicolumn{1}{c}{A$_{2g}$}&\multicolumn{1}{c}{A$_{u}$}\\ \hline
    \LNO &303  &259 &&216\it i &123\it i &158\it i\\
    \LTO &163  &129 &&178\it i &116\it i &1\it i  \\
    \ZSO &255  &241 &&93\it i  &30\it i  &47\it i \\
    \FTO &1014 &735 &&195\it i &75\it i  &---\\
    \MTO &550  &468 &&177\it i &73\it i  &114\it i\\
    \end{tabular}\label{tab:midpoint}
\end{ruledtabular}
\end{center}
\end{table}

The energy differences between R$\bar3$c and R$\bar3$ structures can be
explained by comparing their unstable phonons, for which the frequencies
are listed in Table~\ref{tab:midpoint}. At R$\bar3$c symmetry, all the
LNO-type candidates have {\it two} unstable modes along the rhombohedral
axis direction, namely one A$_{2u}$ and one A$_{2g}$ mode. The A$_{2u}$
mode is IR-active, and it describes the synchronous movement of $A$
cations. The non-polar A$_{2g}$ mode, however, is related to the
out-of-phase movement of the two $A$ cations. Comparing the unstable
modes in the R$\bar3$c and R$\bar3$ structures, we find that the
unstable non-polar mode is absent in the R$\bar3$ structure. Therefore,
the unstable A$_{2g}$ mode is responsible for the energy reduction in
going from the R$\bar3$c to the R$\bar3$ structure. In addition, we
find an unstable E$_u$ mode in \LNO and \FTO\ for both the R$\bar3$c and
R$\bar3$ structures. As the three-fold rotational symmetry is preserved
in our calculation, the E$_u$ modes are not allowed to relax and
further lower the energy.

For the ordered-LNO materials, since the two $A$ cations are not
related by any symmetry even in the ground state, the two $A$ cations
move sequentially so that $\xi_1\ne\xi_2$. Therefore, the reversal path
adopts the R3 symmetry, except at the R$\bar{3}$ midpoint. The only
exception in our calculations is the case of the $udu$ magnetic state
in \MFWO\ and \MWO, where the magnetic moments break inversion symmetry
so that the midpoint structure slightly deviates from R$\bar{3}$ to R3.
Leaving aside this small distortion, the midpoint structures of the
LNO-type and the ordered-LNO paths have the same structural symmetry,
even though the ordered-LNO compounds have lower symmetry in their
ground state.

\begin{figure}
  \includegraphics[width=7.5cm]{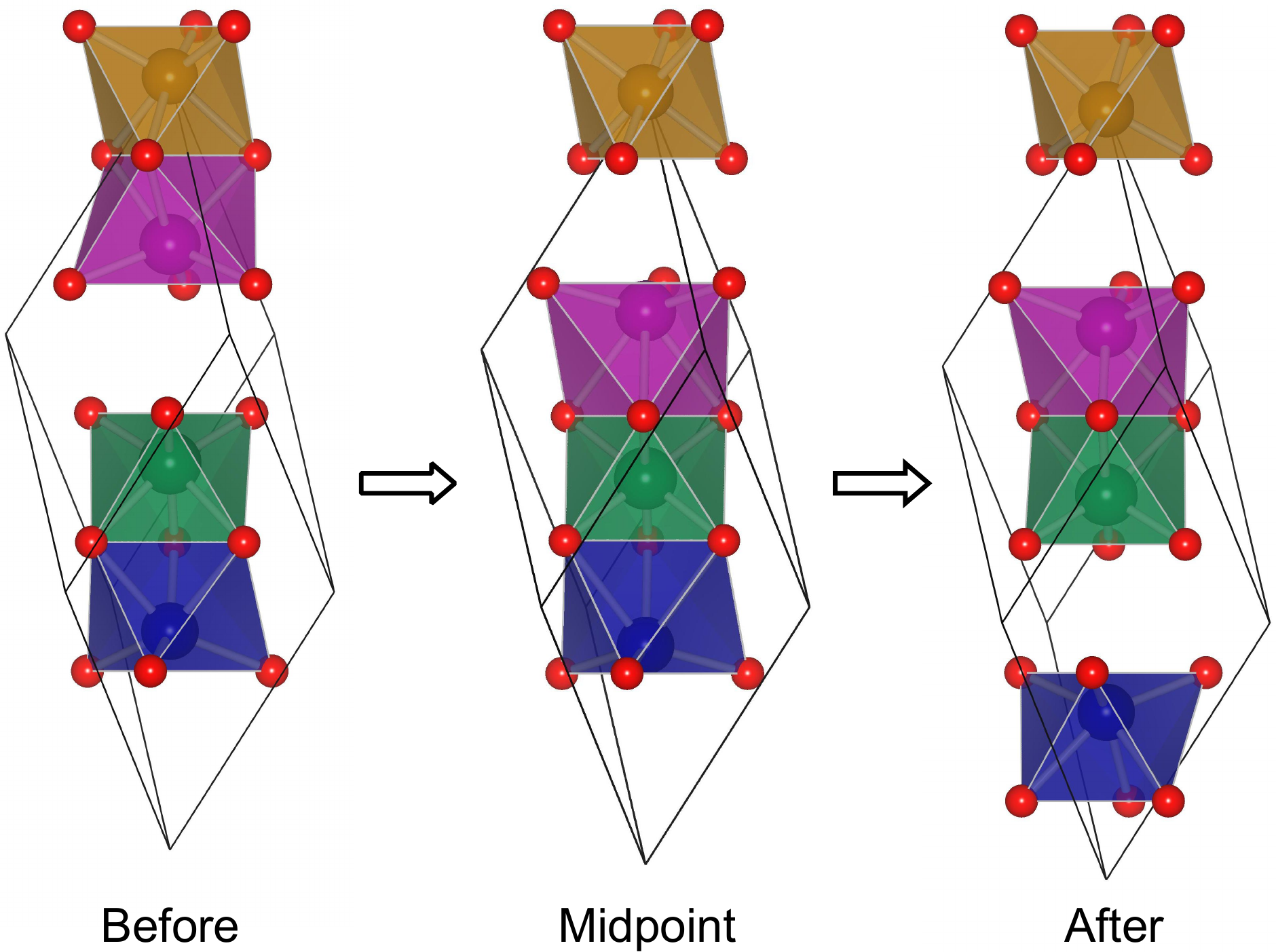}
    \caption{Structural evolution along the polarization reversal path
    of the LNO-type and the ordered-LNO corundum derivatives.
    ``Before" and ``After" are the initial and final structures on the
    reversal path with symmetry R3c for the LNO-type and R3 for the
    ordered-LNO corundum derivatives; ``Midpoint" denotes the
    structure halfway between these
    and exhibits R$\bar{3}$ structural symmetry in both cases.}
    \label{fig:StruEvol}
\end{figure}
\begin{figure}
  \includegraphics[width=7.5cm]{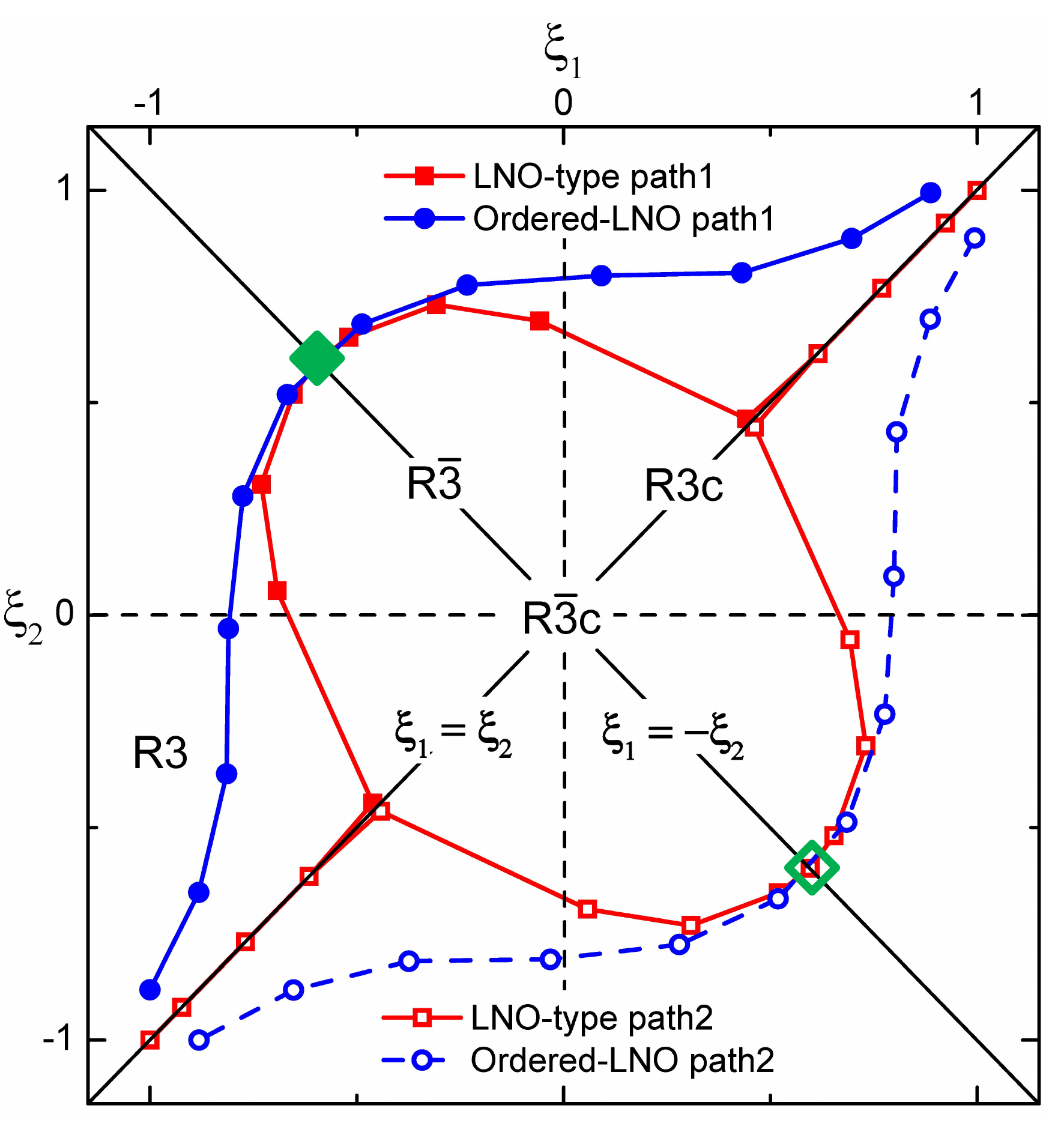}
    \caption{Movements of $A$ cations in LNO-type (red, here \LNO) and
    ordered-LNO (blue, here \MFWO) corundum derivatives along the
    polarization reversal path.
    $\xi_1$ and $\xi_2$ are the distances from $A$ atoms to the oxygen
    planes that are penetrated during the polarization reversal,
    here rescaled to a range between $-1$ and 1.
    The symmetry at an arbitrary $(\xi_1,\xi_2)$ point is R3;
    on the $\xi_1\!=\xi_2$ and $\xi_1\!=\!-\xi_2$ diagonals it is
    raised to R3c and R$\bar{3}$ respectively;
    and at the origin ($\xi_1\!=\xi_2\!=\!0$) it reaches R$\bar{3}$c.
    Green diamonds denote the midpoint structure in the parameter space.
    In the LNO-type case ``path1" and ``path2'' (filled and open
    red square symbols) are equivalent and equally probable, while the
    ordered-LNO system deterministically follows ``path1" (full blue line),
    which becomes ``path2" (dashed blue) under a relabeling
    $\xi_1\leftrightarrow\xi_2$.}
  \label{fig:pathAatom}
\end{figure}
The sequence of movements of the $A$ cations along the FE reversal path
is illustrated in Fig~\ref{fig:StruEvol}, and described quantitatively
using our computed results for \LNO\ and \MFWO\ as paradigmatic examples
in Fig.~\ref{fig:pathAatom}. The ``Before" and ``After'' structures in
Fig.~\ref{fig:StruEvol} correspond to the points at the top right and
bottom left corners of Fig.~\ref{fig:pathAatom} respectively. For the
LNO-type case, the ideal R$\bar{3}$c ``Midpoint" structure would correspond
to the origin on the plot, but the reversal path does not pass through
this point because of an unstable $A_{2g}$ mode along the $\xi_1\!=-\xi_2$
direction. The ``bubble" in the center confirms the significant effect
of the unstable $A_{2g}$ mode. Our ``Midpoint'' in Fig.~\ref{fig:StruEvol}
is thus displaced from the origin along the line $\xi_1\!=\!-\xi_2$.
There is a spontaneous breaking of symmetry at the point where the
structure departs from the $\xi_1\!=\xi_2$ diagonal; at this point
the system ``randomly'' makes a choice between two equivalent paths,
marked by filled and open red symbols in Fig.~\ref{fig:pathAatom}.

For the ordered-LNO materials the two $A$ cations are inequivalent,
and one of them is already closer to the oxygen plane in the ground
state. Let this be the one labeled by $\xi_1$. It is energetically
favorable for this particular $A$ cation to migrate first in the
reversal path, which causes either the $B$ or $B'$ cation to be
sandwiched between two $A$ cations in the `Midpoint'' structure as
illustrated in Fig.~\ref{fig:StruEvol}.
The system thus deterministically follows the path indicated by the
full blue line in Fig.~\ref{fig:pathAatom}, with the configuration
always staying on one side of the $\xi_1\!=\xi_2$ diagonal. If we
would reverse the convention on the definition of $\xi_1$ and $\xi_2$, the
system would be described by the dashed blue path in Fig.~\ref{fig:pathAatom}.

\subsection{Polarization reversal barrier}

\begin{figure}
  \includegraphics[width=8cm]{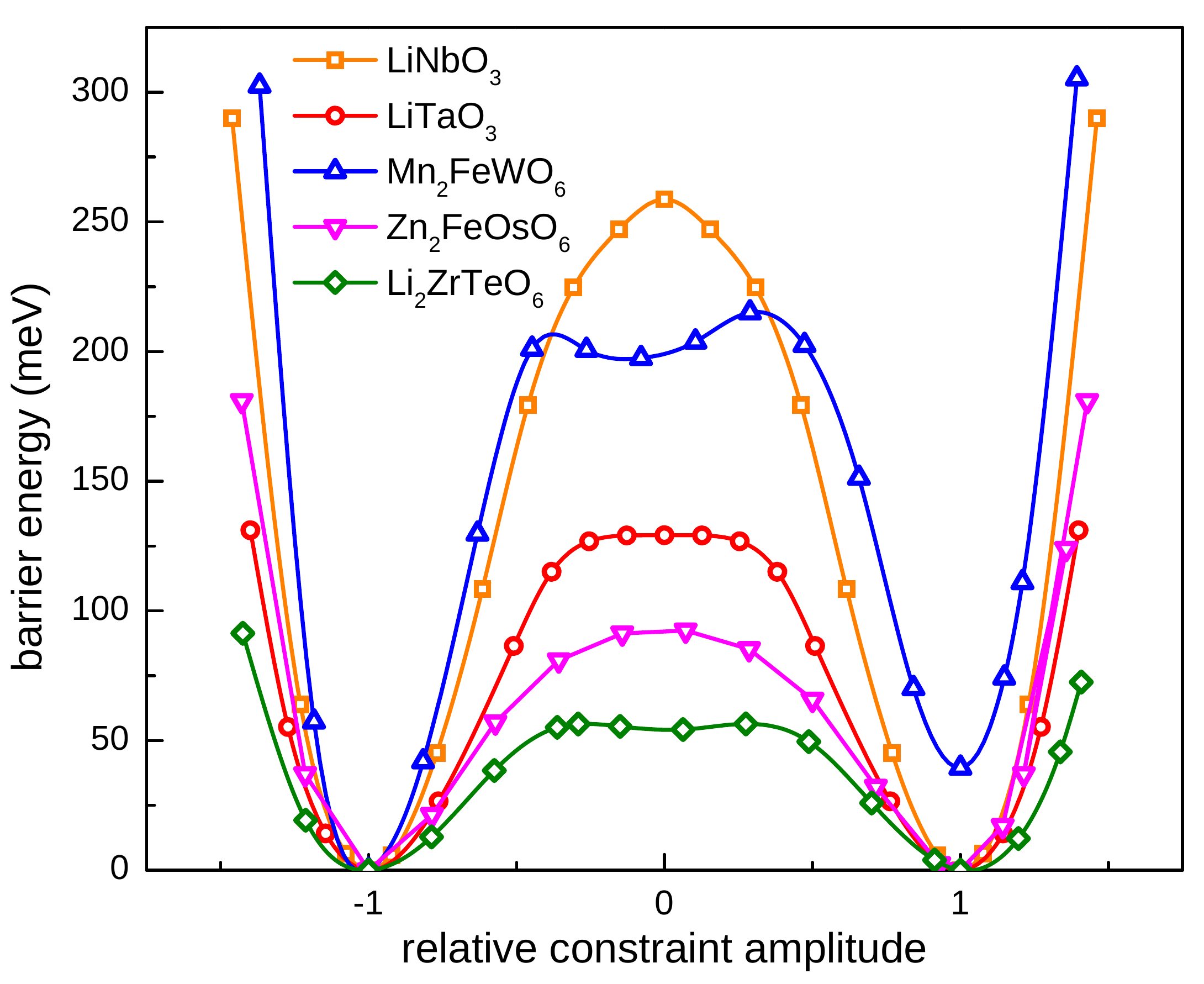}
    \caption{The polarization reversal energy profile for \LNO, \LTO,
    \MFWO, \ZFOO, and \LZTO. }
    \label{fig:barrier}
\end{figure}

Using the methods described in Secs.~\ref{sec:profile} and \ref{sec:fpm},
we compute the relaxed structures and energies for a sequence of
constrained values of our ``reaction coordinate'' $\xi\!=\xi_1+\xi_2$
for each material of interest. A selection of results for the energy
along the path are presented in Fig.~\ref{fig:barrier}, and quantitative
results for the energy barrier $E_{\rm {barrier}}$ and the spontaneous
polarization $P_{\rm S}$ in the ground-state structure are reported in
Table~\ref{tab:barrier}.

\begin{table}[b]
\begin{center}
\begin{ruledtabular}
    \caption{Coherent polarization reversal barrier $E_{\rm {barrier}}$
    (meV) per unit cell and spontaneous polarization $P_{\rm S}$
    ($\mu$C/cm$^2$) for FE candidates.}
    \begin{tabular}{cccccccc}
    \multicolumn{1}{c}{LNO-type} &\multicolumn{1}{c}{$E_{\rm {barrier}}$  } &\multicolumn{1}{c}{$P_{\rm S}$ }&&
    \multicolumn{1}{c}{Ordered-LNO} &\multicolumn{1}{c}{$E_{\rm {barrier}}$ } &\multicolumn{1}{c}{$P_{\rm S}$ }\\ \hline
    \LNO &259 & 82 && \LZTO     & 57 & 33 \\
    \LTO &129 & 57 && \LHTO     & 61 & 32 \\
    \ZSO &241 & 57 && \MFWO     &215 & 63 \\
    \FTO &763 & 105&&$uud$ \MWO &240 & 69 \\
    \MTO &468 & 94 &&$udu$ \MWO &272 & 70 \\
          &   &    && \ZFOO     & 92 & 52 \\
    \end{tabular}\label{tab:barrier}
\end{ruledtabular}
\end{center}
\end{table}

We find that the cations that are sandwiched in the midpoint structures
of ordered-LNO candidates are Zr for \LZTO, Hf for \LHTO, W for \MFWO\
and \MWO, and Os for \ZFOO. These results are consistent with the
analysis of $\xi_1$ and $\xi_2$ in the ground state, as mentioned in
Sec.~\ref{sec:sympath}. We find that the energy difference between
the $B$ and $B'$ sandwiched midpoint structures can be attributed mainly
to the Madelung energy, as shown in the Supplement. Among the computed energy
barriers, those for \ZSO, \LZTO, \LHTO, \MFWO, \MWO, and \ZFOO\ are
lower than or comparable to those of the established FE materials \LNO\
and \LTO.

\begin{figure}
  \includegraphics[width=8cm]{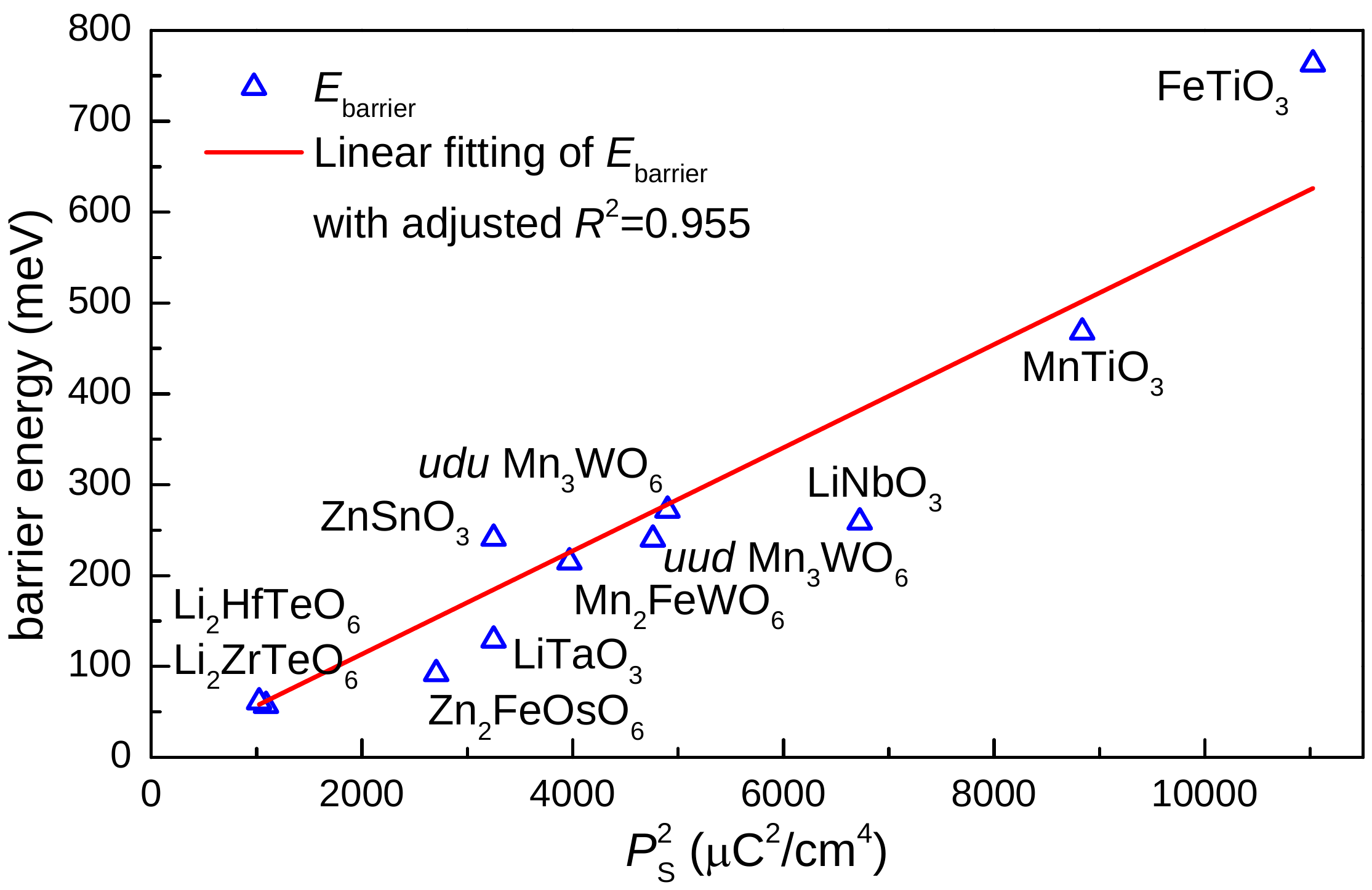}
    \caption{Empirical proportionality between the coherent FE energy
    barrier and $P^2_{\rm S}$. The red curve is the fitting polynomial
    $E_{\rm {barrier}}\!=(\mu/2)P_{\rm S}^2$, with $\mu/2\!=0.057$.}
    \label{fig:dipoleE}
\end{figure}

We have analyzed our calculations in an attempt to extract empirical
rules of thumb that may help point in the direction of more new
materials with low reversal barriers. Firstly, we have considered how
the energy barriers are correlated with the spontaneous polarizations.
In a FE material the energy $E$ is often approximated as a double well
of the form $E(P)\!=E_0-\mu P^2+\nu P^4$ with positive $\mu$ and $\nu$.
Minimizing $E(P)$ within this model gives the spontaneous polarization
as $P_{\rm S}^2\!=\mu/2\nu$ and the
energy barrier $E_{\rm {barrier}}\!=E(0)-E(P_{\rm S})\!=\mu^2/4\nu$,
which can also be written as $E_{\rm {barrier}}\!=(\mu/2)P_{\rm S}^2$.
Thus, as long as $\mu$ can be taken as approximately constant,
$E_{\rm {barrier}}$ is proportional to $P_{\rm S}^2$. Interestingly,
we find that our computed coherent barrier energies $E_{\rm {barrier}}$
roughly follow this trend with $\mu/2\!=0.057$ as shown in Fig.~\ref{fig:dipoleE}.
Therefore we suggest that FE corundum derivatives are more likely to
be discovered in materials having a relatively low spontaneous
polarization.

\begin{figure}
  \includegraphics[width=8cm]{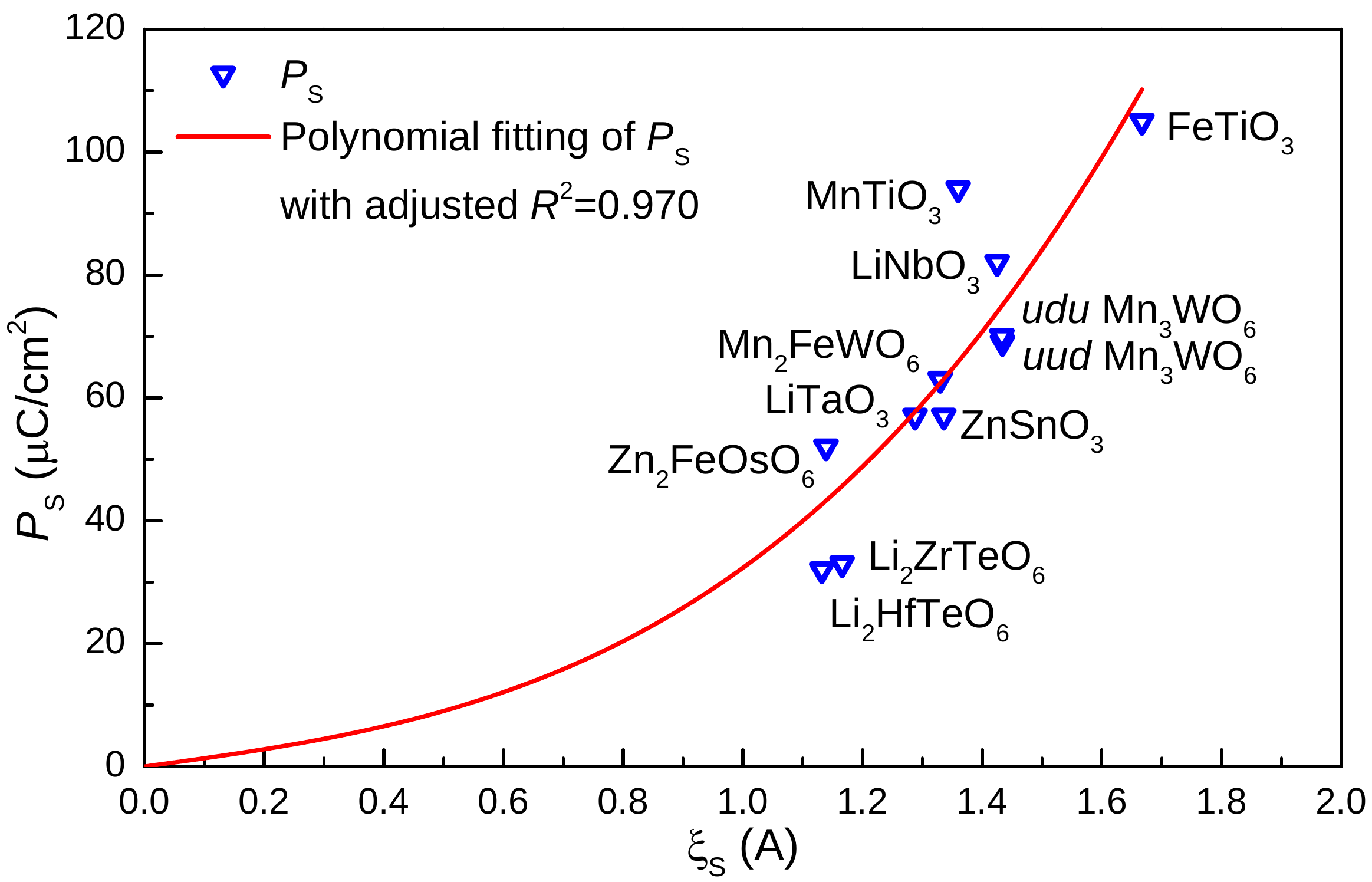}
    \caption{Empirical correlation between the spontaneous polarization
    and the reaction coordinate $\xi$ in the ground state. The red curve
    is the fitting polynomial $P_{\rm S}\!=m\xi_{\rm S}+n\xi_{\rm S}^3$
    with $m\!=13.3$ and $n\!=19.0$.}
    \label{fig:PvsXi}
\end{figure}

Furthermore, we have investigated the correlation between the
spontaneous polarizations and the geometric properties of the crystals.
Our results suggest that for each FE candidate, the polarization $P$
along the reversal path is monotonically related to the reaction
coordinate $\xi$, with
the approximate relationship $P(\xi)\!=m\xi+n\xi^3$. The parameters
$m$ and $n$ are different in each material, and they are determined by
several factors that are not included in the reaction coordinate $\xi$,
such as the displacements of the $B$ cations and the valence states of
the $A$ cations. Despite these differences between materials, we find
that the spontaneous polarizations $P_{\rm S}$ of corundum derivatives
are approximately related to the reaction coordinate $\xi_{\rm S}$
in the spontaneously polarized ground state by a corresponding formula
$P_{\rm S}\!=m\xi_{\rm S}+n\xi_{\rm S}^3$ with $m\!=13.3$ and $n\!=19.0$
as shown in Fig.~\ref{fig:PvsXi}. As the distance between $A$ cations
and oxygen planes can be experimentally determined, this empirical rule
can provide a rough estimation of the spontaneous polarization.

\subsection{Insulating vs.~conducting}
Our density-of-states calculations (not shown) indicate that \FTO\ and
\MFWO\ are conducting along the central portion of the polarization
reversal path. A detailed analysis of the occupied $d$ orbitals along
the path reveals the reason for this metal-insulator transition. In
the local octahedral environment of the ground state, the $d$ orbitals
are split into triply degenerated $t_{2g}$ and doubly degenerated
$e_g$ orbitals. Under the threefold rotational symmetry, the $t_{2g}$
orbitals are further split into $a_{1g}$ and doubly degenerate $e_g'$
irreps. The $a_{1g}$ state has orbital character d$_{z^2}$ with charge
lobes directed along the rhombohedral axis, and since these lie closer
to the neighboring cations, the energy of the $a_{1g}$ state is lowered.
Therefore, the ground-state arrangement of $d$ orbitals in order of
increasing energy is $a_{1g}$ followed by $e_g'$ and then $e_g$.
In \FTO\ and \MFWO, Fe is in the 2+ valence state and has a $d^6$
configuration. In the ground state, five electrons fully occupy one
spin channel and the remaining one occupies the $a_{1g}$ orbital in the
minority spin channel. However, during the polarization reversal
process, the Fe$^{2+}$ ion temporarily moves away from its neighboring
cations, and as a result, the $a_{1g}$ orbital is no longer
energetically favored. Instead, the minority electron occupies the
doubly degenerate $e_g'$ orbitals, leading to a metallic state.
Since a metallic state along the polarization reversal path could short
out the applied bias, it may be that the switching of polarization is
not possible in such cases. We propose that $d^3$, $d^5$, and $d^8$
orbital configurations should be much more likely to avoid this
conducting problem, and are therefore more suitable targets in the
search for ferroelectrically switchable magnetic corundum derivatives.

\section{Summary}

In this paper, we have proposed a method to study the coherent FE reversal
path of the corundum derivative family. By analyzing the structures,
we have shown that only the LNO-type and the ordered-LNO corundum
derivatives can be FE in the usual sense.
We have calculated the energy profiles of the
reversal paths using first-principles density-functional methods.
Our calculations reveal that the symmetry of
the FE barrier structure is lower than that of the
paraelectric phase. According to our calculations, \ZSO, \LZTO, \LHTO, \MWO,
and \ZFOO\ are predicted to be possible new FE materials. We have
found empirically that the energy barrier is roughly proportional to
the the square of the spontaneous
polarization, and that the spontaneous polarization is strongly
correlated with the
reaction coordinate $\xi$ in the ground state. Finally, we
have also argued that magnetic corundum derivatives are unlikely
to be suitable for FE switching unless the magnetic ion is
$d^3$, $d^5$ or $d^8$, since metallic configurations otherwise tend to
appear along the FE reversal path.

\section{Acknowledgment}


\parbox{8.1cm}{
We thank Martha Greenblatt and Manrong Li for many useful discussions.
The work was supported by ONR grant N00014-12-1-1035.
}\\



\begin{thebibliography}{99}

\bibitem{BookRabe} Rabe, Karin M., Charles H. Ahn, Jean-Marc Triscone. \textit{Physics of Ferroelectrics}. Springer, 2008. Print.

\bibitem{BookScott} Scott, James F.. \textit{Ferroelectric Memories}. Springer, 2000. Print.

\bibitem{Scott06} W. Eerenstein, N. D. Mathur and J. F. Scott, Nature {\bf 442}, 759 (2006).

\bibitem{Cheong&Mostovoy07} S.-W. Cheong and M. Mostovoy, Nat. Mater. {\bf 6}, 13 (2007).

\bibitem{Ramesh&Spaldin07} R. Ramesh and N. A. Spaldin, Nat. Mater. {\bf 6}, 21 (2007).

\bibitem{Wang09} K. F. Wang, J.-M. Liu and Z. F. Ren, Adv. Phys. {\bf 58}, 321 (2009).

\bibitem{Khomskii09} D. Khomskii, Physics {\bf 2}, 20 (2009).

\bibitem{Picozzi15} P. Barone, S. Picozzi, C. R. Physique {\bf 16}, 143 (2015).

\bibitem{Ghosez15} J. Varignon, N. C. Bristowe, \'{E}. Bousquet, P. Ghosez, C. R. Physique {\bf 16}, 153 (2015).

\bibitem{pvskFE92} R. E. Cohen, Nature {\bf 358}, 136 (1992).

\bibitem{pvskFE97} D. Vanderbilt, Curr. Opin. Solid State Mater. Sci. {\bf 2}, 701 (1997).

\bibitem{dpvskFE13} A. T. Mulder, N. A. Benedek, J. M. Rondinelli, and C. J. Fennie, Adv. Funct. Mater {\bf 23}, 4810 (2013).

\bibitem{dpvskMF14} H. J. Zhao, W. Ren, Y. Yang, J. \'{I}\~{n}\.{i}guez, X. M. Chen, and L. Bellaiche, Nat. Commun. {\bf 5}, 4021 (2014).

\bibitem{LNO/LTO} A. M. Glass and M. E. Lines, Phys. Rev. B {\bf 13}, 180 (1976).

\bibitem{LNO/LTOcal} I. Inbar and R. E. Cohen, Phys. Rev. B {\bf 53}, 1193 (1996).

\bibitem{LNO/LTOcal2} M. Veithen and Ph. Ghosez, Phys. Rev. B {\bf 65}, 214302 (2002).

\bibitem{CuNb/TaO} A. W. Sleight and C. T. Prewitt, Mat. Res. Bull. {\bf 5}, 207 (1970).

\bibitem{Fe/Mn/NiTiO} C. J. Fennie, Phys. Rev. Lett. {\bf 100}, 167203 (2008).

\bibitem{FeTiO} T. Varga, A. Kumar, E.Vlahos, S. Denev, M. Park, S. Hong, T. Sanehira, Y. Wang, C. J. Fennie, S. K. Streiffer, X. Ke, P. Schiffer, V. Gopalan, and J. F. Mitchell, Phys. Rev. Lett. {\bf 103}, 047601 (2009).

\bibitem{ZnSnOFE} J. Y. Son, G. Lee, M.-H. Jo, H. Kim, H. M. Jang, and Y.-H. Shin, J. Am. Chem. Soc. {\bf 131}, 8386 (2009).

\bibitem{CdPbO/PbNiO} Y. Inaguma, M. Yoshida, T. Tsuchiya, A. Aimi, K. Tanaka, T. Katsumata, D. Mori, J. Phys. Conf. Ser. {\bf 215}, 012131 (2010).

\bibitem{MnTi/SnO} A. Aimi, T. Katsumata, D. Mori, D. Fu, M. Itoh, T. Ky\^{o}men, K. Hiraki, T. Takahashi, and Y. Inaguma, Inorg. Chem. {\bf 50}, 6392 (2011)

\bibitem{PbNiO} X. F. Hao, A. Stroppa, S. Picozzi, A. Filippetti, and C. Franchini, Phys. Rev. B {\bf 86}, 014116 (2012).

\bibitem{ScFeO} T. Kawamoto, K. Fujita, I. Yamada, T. Matoba, S. J. Kim, P. Gao, X. Pan, S. D. Findlay, C. Tassel, H. Kageyama, A. J. Studer, J. Hester, T. Irifune, H. Akamatsu, and K. Tanaka, J. Am. Chem. Soc. {\bf 136}, 15291 (2014).

\bibitem{LiZr/HfTeO} J. Choisnet, A. Rulmont, P.Tarte, J. Solid State Chem. {\bf 75}, 124 (1988).

\bibitem{Mn2FeSbO} R. Mathieu, S. A. Ivanov, G. V. Bazuev, M. Hudl, P. Lazor, I.V. Solovyev, P. Nordblad, Appl. Phys. Lett. {\bf 98}, 202505 (2011).

\bibitem{Mn2FeNb/TaO} M.-R. Li, D. Walker, M. Retuerto, T. Sarkar, J. Hadermann, P. W. Stephens, M. Croft, A. Ignatov, C. P. Grams, J. Hemberger, I. Nowik, P. S. Halasyamani, T. T. Tran, S. Mukherjee, T. S. Dasgupta, and M. Greenblatt, Angew. Chem. Int. Ed. {\bf 52}, 8406 (2013).

\bibitem{Mn2FeMoO} M.-R. Li, M. Retuerto, D. Walker, T. Sarkar, P. W. Stephens, S. Mukherjee, T. S. Dasgupta, J. P. Hodges, M. Croft, C. P. Grams, J. Hemberger, J. S\'{a}nchez-Ben\'{\i}tez, A.Huq, F. O. Saouma, J. I. Jang, and M. Greenblatt, Angew. Chem. Int. Ed. {\bf 53}, 10774 (2014).

\bibitem{Zn2FeTaO} M.-R. Li, P. W. Stephens, M. Retuerto, T. Sarkar, C. P. Grams, J. Hemberger, M. C. Croft, D. Walker, and M. Greenblatt, J. Am. Chem. Soc. {\bf 136}, 8508 (2014).

\bibitem{Mn2FeWO} M.-R. Li, M. Croft, P. W. Stephens, M. Ye, D. Vanderbilt, M. Retuerto, Z. Deng, C. P. Grams, J. Hemberger, J. Hadermann, W.-M. Li, C.-Q. Jin, F. O. Saouma, J. I. Jang, H. Akamatsu, V. Gopalan, D. Walker, and M. Greenblatt, Adv. Mater. {\bf 27}, 2177 (2015).

\bibitem{Zn2FeOsO} P. S. Wang, W. Ren, L. Bellaiche, and H. J. Xiang, Phys. Rev. Lett. {\bf 114}, 147204 (2015).

\bibitem{Mn3WO} Private communication with Prof. Martha Greenblatt and Dr. Manrong Li.

\bibitem{VASP} G. Kresse and J. Furthm\"{u}ller, Phys. Rev. B {\bf 54}, 11169 (1996).

\bibitem{PBEsol} J.P. Perdew, A. Ruzsinszky, G.I. Csonka, O.A. Vydrov, G.E. Scuseria, L.A. Constantin, X. Zhou, and K. Burke, Phys. Rev. Lett. {\bf 100}, 136406 (2008).

\bibitem{PAW} P. E. Blochl, Phys. Rev. B {\bf 50}, 17953(1994); G. Kresse and D. Joubert, Phys. Rev. B {\bf 59}, 1758 (1999).

\bibitem{DFTU} S. L. Dudarev, G. A. Botton, S. Y. Savrasov, C. J. Humphreys and A. P. Sutton, Phys. Rev. B {\bf 57}, 1505 (1998).

\bibitem{ModernPolar} R. D. King-Smith and D. Vanderbilt, Phys. Rev. B {\bf 47}, 1651 (1993).

\end{thebibliography}
\end{document}



\title{Supplement: Ferroelectricity in corundum derivatives}

\author{Meng Ye}
\email{mengye@physics.rutgers.edu}
\affiliation{
Department of Physics \& Astronomy, Rutgers University,
Piscataway, New Jersey 08854, USA}

\author{David Vanderbilt}
\affiliation{
Department of Physics \& Astronomy, Rutgers University,
Piscataway, New Jersey 08854, USA}

\date{\today}

\begin{abstract}
\end{abstract}

\maketitle

The properties of FE materials are sensitive to atomic displacements
and strain, so it is essential to start our calculation with accurate
structural parameters. The lattice constants and Wyckoff positions
obtained from our calculations are summarized in the Tables~\ref{tab:StruABO}
and \ref{tab:StruABB'}, with experimental results provided for
reference. Our structure parameters are very close to the experimental
results.

To investigate the magnetic structures of \MFWO\ and \MWO, four different
types of unit cell are considered in the calculation. We adopt a notation
like ``$udu$" to describe the possible spin structure, where ``$u$" is
spin-up, ``$d$" is spin-down, and the spins are given on atom $\beta$,
$\delta$ and $\gamma$, in that order. The four possible states (not
counting those that are trivially related by a global spin reversal)
are $uuu$, $uud$, $udu$, and $udd$. The energy of each fully-relaxed
magnetic structure is listed in Table~\ref{tab:magGS}. Of those, the
most stable state for both \MFWO\ and \MWO\ is $udu$. However, for \MWO,
the energy difference between the $uud$ and $udu$ states is tiny, so
we considered the polarization reversal for both magnetic states.

The magnetic ground states of \MFWO\ and \MWO\ can be understood by
analyzing the super-exchange interactions between $A_1$, $A_2$ and
$B$ cations (also shown as $\beta$, $\gamma$ and $\delta$ cations).
The magnetic moments are
coupled through the oxygen octahedra, and there are three independent
coupling constants. The moments on $A_2$ and $B$ sites are coupled through
face-sharing and corner-sharing oxygen octahedra with strength $J_{A_2B}$;
the $A_1$ and $B$ moments are coupled through edge-sharing octahedra
with strength $J_{A_1B}$; and the $A_1$ and $A_2$ moments are coupled
through corner-sharing octahedra with strength $J_{A_1A_2}$. Then the
magnetic energy $E_{\rm{mag}}$ can be written as
\beq
E_{\rm{mag}} = J_{A_1B} \hat S_{A_1}\cdot \hat S_B +
J_{A_2B} \hat S_{A_2}\cdot \hat S_B + J_{A_1A_2} \hat S_{A_1}\cdot \hat S_{A_2}\,,
\label{eq:magGS}
\eeq
where $\hat S$ represent the spin on each site. Substituting the
energy of different magnetic orderings in Table \ref{tab:magGS}
into Eq.(\ref{eq:magGS}), we find that the coupling constants are all
positive. This result implies that the three magnetic moments all favor
AFM coupling. However, it is impossible to make three collinear spins
couple antiferromagnetically, and this frustration implies that one
pair must be
ferromagnetically coupled.

In Table \ref{tab:EX} we list the relative spin orientations of the magnetic ions.
Since the face-sharing coupling $J_{A_2B}$ is the strongest, it is not surprising
that the $A_2$ and $B$ moments couple antiferromagnetically; the competition
between $J_{A_1B}$ and $J_{A_1A_2}$ then determines the magnetic ground
state. In \MWO, these two couplings are comparable, so the energy
difference between the $uud$ and $udu$ states is tiny. In the case of \MFWO,
the magnetic moment on the $B$ cation is smaller, so the coupling $J_{A_1B}$
is weaker than $J_{A_1A_2}$. Therefore, the lowest-energy state is $udu$.

For the ordered-LNO materials the two $A$ cations are inequivalent,
and one of them is already closer to the oxygen plane in the ground
state ($\xi_{1 \rm S}\!\neq\xi_{2 \rm S}$).  Therefore it is
energetically favorable for this particular $A$ cation to migrate
first in the reversal path, which causes either the $B$ or $B'$ cation
to be sandwiched between two $A$ cations in the midpoint structure.
If $\xi_{1 \rm S}>\xi_{2 \rm S}$, the sandwiched cation is $B$;
otherwise it is $B'$. By calculating the energy
difference $\Delta E\!=E_B-E_{B'}$ between the $B$ and $B'$ sandwiched
midpoint structures, we find that the cations that are sandwiched in
the midpoint structures are Zr for \LZTO, Hf for \LHTO, W for \MFWO\
and \MWO, and Os for \ZFOO. These results are consistent with the
analysis of $\xi_{1 \rm S}$ and $\xi_{2 \rm S}$, as shown in
Table~\ref{tab:sandwich}. We find that the energy difference between
the two distinct midpoint structures is mainly due to the difference
$\Delta E^{\rm M}\!=E_B^{\rm M}-E_{B'}^{\rm M}$ between the Madelung
energies of the two structures in a simple point-ion model, as
shown in Table.~\ref{tab:sandwich}.


\begin{table}
\begin{center}
\begin{ruledtabular}
    \caption{Rhombohedral structural parameters of LNO-type \ABO\
    corundum derivatives \LNO, \LTO, \ZSO, \FTO\ and \MTO\ from
    first-principles calculations and experiments.
    The Wyckoff positions are 2a for $A$ and $B$ cations, and 6c for
    oxygen anions (note that $A_x\!=A_y\!=A_z$ and $B_x\!=B_y\!=B_z$).
    The origin is defined by setting the Wyckoff position $B_x$ to zero.}
    \begin{tabular}{ccccccccc}
    &&\multicolumn{2}{c}{Lattice constants} &\multicolumn{5}{c}{Wyckoff positions}\\
    &&\multicolumn{1}{c}{a ({\AA})} &\multicolumn{1}{c}{$\alpha$ ($^\circ$)} &\multicolumn{1}{c}{$A_x$} &\multicolumn{1}{c}{$B_x$}
    &\multicolumn{1}{c}{O$_x$} &\multicolumn{1}{c}{O$_y$} &\multicolumn{1}{c}{O$_z$}\\\hline
    \multirow{2}{*}{\LNO} &Calc.              &5.486 &56.0 &0.282 &0.000 &0.360 &0.719 &0.112 \\
                          &Exp. \cite{LNOlatt}&5.494 &55.9 &0.280 &0.000 &0.359 &0.720 &0.111 \\ \hline
    \multirow{2}{*}{\LTO} &Calc.              &5.476 &56.2 &0.284 &0.000 &0.365 &0.726 &0.119 \\
                          &Exp. \cite{LTOlatt}&5.473 &56.2 &0.291 &0.000 &0.368 &0.732 &0.124 \\ \hline
    \multirow{2}{*}{\ZSO} &Calc.              &5.584 &56.5 &0.283 &0.000 &0.392 &0.709 &0.104 \\
                          &Exp. \cite{ZSOlatt}&5.569 &56.4 &0.286 &0.000 &0.381 &0.721 &0.111 \\ \hline
    \multirow{2}{*}{\FTO} &Calc.              &5.434 &56.5 &0.290 &0.000 &0.364 &0.721 &0.104 \\
                          &Exp. \cite{FTOlatt}&5.458 &56.0 &0.287 &0.000 &0.364 &0.720 &0.109 \\ \hline
    \multirow{2}{*}{\MTO} &Calc.              &5.481 &56.6 &0.279 &0.000 &0.348 &0.721 &0.120 \\
                          &Exp.\cite{MTOlatt} &5.455 &56.8 &0.276 &0.000 &0.345 &0.731 &0.128 \\
    \end{tabular}\label{tab:StruABO}
\end{ruledtabular}
\end{center}
\end{table}
\begin{table}
\begin{center}
\begin{ruledtabular}
    \caption{Rhombohedral structure parameters of ordered-LNO \ABB'O
    corundum derivatives \LZTO, \LHTO, \MFWO, \MWO\ and \ZFOO\ from
    first-principles calculations and experiments.
    Wyckoff positions are 1a for $A_1$, $A_2$, $B$ and $B'$ cations,
    and 3b for O$_1$ and O$_2$ anions.
    The origin is defined by setting the Wyckoff position $B'_x$ to zero.
    For ordered-LNO \MWO\ and \ZFOO, no experimental results are available.}
    \begin{tabular}{lcccccccccccccc}
    &&\multicolumn{1}{c}{Magnetic} &\multicolumn{2}{c}{Lattice constants} &\multicolumn{10}{c}{Wyckoff position}\\
    &&\multicolumn{1}{c}{order}&\multicolumn{1}{c}{a ({\AA})} &\multicolumn{1}{c}{$\alpha$ ($^\circ$)}
    &\multicolumn{1}{c}{$A_{1x}$} &\multicolumn{1}{c}{$A_{2x}$} &\multicolumn{1}{c}{$B_x$} &\multicolumn{1}{c}{$B'_x$}
    &\multicolumn{1}{c}{O$_{1x}$} &\multicolumn{1}{c}{O$_{1y}$} &\multicolumn{1}{c}{O$_{1z}$}
    &\multicolumn{1}{c}{O$_{2x}$} &\multicolumn{1}{c}{O$_{2y}$} &\multicolumn{1}{c}{O$_{2z}$}\\ \hline
    \multirow{2}{*}{\LZTO}&Calc.                 &---   &5.526 &56.1 &0.291 &0.781 &0.504 &0.000 &0.366 &0.745 &0.111 &0.628 &0.219 &0.895\\
                          &Exp. \cite{LiZr/HfTeO}&---   &5.497 &56.1 &0.298 &0.768 &0.507 &0.000 &0.390 &0.729 &0.133 &0.621 &0.235 &0.893\\\hline
    \multirow{2}{*}{\LHTO}&Calc.                 &---   &5.480 &56.3 &0.289 &0.781 &0.504 &0.000 &0.366 &0.743 &0.115 &0.629 &0.222 &0.889\\
                          &Exp.                  &---   & NA   & NA  & NA   & NA   & NA   & NA   & NA   &  NA  &  NA  &  NA  &  NA  &  NA \\\hline
    \multirow{2}{*}{\MFWO}&Calc.                 &$udu$ &5.531 &57.3 &0.286 &0.787 &0.506 &0.000 &0.356 &0.744 &0.109 &0.632 &0.207 &0.884\\
                          &Exp. \cite{Mn2FeWO}   & NA   &5.562 &56.9 &0.278 &0.785 &0.493 &0.000 &0.347 &0.745 &0.102 &0.631 &0.197 &0.885\\\hline
    \multirow{3}{*}{\MWO} &Calc.                 &$udu$ &5.607 &56.7 &0.287 &0.790 &0.508 &0.000 &0.356 &0.742 &0.102 &0.631 &0.199 &0.893\\
                          &Calc.                 &$uud$ &5.613 &56.6 &0.283 &0.779 &0.493 &0.000 &0.384 &0.693 &0.124 &0.595 &0.234 &0.848\\
                          &Exp. \cite{Mn3WO}     & NA   &5.605 &56.7 &0.283 &0.788 &0.510 &0.000 &0.351 &0.744 &0.104 &0.631 &0.190 &0.901\\\hline
    \multirow{2}{*}{\ZFOO}&Calc.&FiM   &5.410 &56.7 &0.284 &0.783 &0.504 &0.000 &0.376 &0.732 &0.114 &0.619 &0.223 &0.885\\
                          &Exp.                  & NA   & NA   & NA  & NA   & NA   & NA   & NA   & NA   &  NA  &  NA  &  NA  &  NA  &  NA \\
    \end{tabular}\label{tab:StruABB'}
\end{ruledtabular}
\end{center}
\end{table}

\begin{table}
\begin{center}
\begin{ruledtabular}
    \caption{Magnetic energy of different magnetic states relative to
    the lowest-energy state in \MFWO\ and \MWO, in units of meV per
    formula unit (f.u.).}
    \begin{tabular}{ldddd}
            &\multicolumn{4}{c}{Energy (meV/f.u.)}\\
            &\multicolumn{1}{c}{$uuu$} &\multicolumn{1}{c}{$uud$} &\multicolumn{1}{c}{$udd$} &\multicolumn{1}{c}{$udu$}\\ \hline
    \MFWO   &90.2  &32.1 &39.5  &0.0\\
    \MWO    &101.8 &1.0  &19.2  &0.0\\
    \end{tabular}\label{tab:magGS}
\end{ruledtabular}
\end{center}
\end{table}

\begin{table}
\begin{center}
\begin{ruledtabular}
    \caption{Relative spin direction between different magnetic ions
    in \MFWO\ and \MWO.\hfill}
    \begin{tabular}{lcccc}
    &\multicolumn{1}{c}{$uuu$} &\multicolumn{1}{c}{$uud$} &\multicolumn{1}{c}{$udd$} &\multicolumn{1}{c}{$udu$}\\ \hline
    $A_1$ and $B$   & FM & AFM & FM  & AFM \\
    $A_2$ and $B$   & FM & AFM & AFM & FM  \\
    $A_1$ and $A_2$ & FM & FM  & AFM & AFM \\
    \end{tabular} \label{tab:EX}
\end{ruledtabular}
\end{center}
\end{table}

\begin{table}
\begin{center}
\begin{ruledtabular}
    \caption{Ground state and midpoint structures of ordered-LNO
    candidates and the energy differences between $B$ and $B'$
    sandwiched midpoint structures. The distances between $A$ cations
    and the oxygen planes in the ground state are characterized by
    $\xi_{1 \rm S}$ and $\xi_{2 \rm S}$. The energy difference between
    the $B$ and $B'$ sandwiched midpoint structures is $\Delta E$. The
    Madelung energy difference between the $B$ and $B'$ sandwiched
    midpoint structures is $\Delta E^{\rm M}$. }
    \begin{tabular}{lcccccdd}
    \multicolumn{1}{c}{Ordered-LNO} &\multicolumn{1}{c}{$B$} &\multicolumn{1}{c}{$B'$} &\multicolumn{1}{c}{Sandwiched cation} &\multicolumn{1}{c}{$\xi_{1 \rm S}$ (\AA)} &\multicolumn{1}{c}{$\xi_{2 \rm S}$ (\AA)} &\multicolumn{1}{c}{$\Delta E^{\rm M}$ (meV)} &\multicolumn{1}{c}{$\Delta E$ (meV)}  \\ \hline
    \LZTO       & Zr & Te & Zr &0.607 &0.559 & -186 & -48 \\
    \LHTO       & Hf & Te & Hf &0.582 &0.550 & -94  & -29 \\
    \MFWO       & Fe & W  & W &0.623 &0.707 & 2085 & 225 \\
    $uud$ \MWO  & Mn & W  & W &0.635 &0.799 & 1666 & 266 \\
    $udu$ \MWO  & Mn & W  & W &0.642 &0.790 & 1499 & 290 \\
    \ZFOO       & Fe & Os & Os &0.565 &0.574 & 382  & 102 \\
    \end{tabular}\label{tab:sandwich}
\end{ruledtabular}
\end{center}
\end{table}